# Scattering dominated high-temperature phase of $1T$-TiSe$_2$: an optical conductivity study

K. Velebit,[1,2] P. Popčević,[1,3] I. Batistić,[4] M. Eichler,[2] H. Berger,[5] L.Forró,[5] M. Dressel,[2] N. Barišić,[1,2,3] and E. Tutiš[1,*]

[1]*Institute of Physics, Bijenička c. 46, HR-10000 Zagreb, Croatia*
[2]*1. Physikalisches Institut, Universität Stuttgart, Pfaffenwaldring 57, 70569 Stuttgart, Germany*
[3]*Institute of Solid State Physics, Vienna University of Technology, 1040 Vienna, Austria*
[4]*Department of Physics, Faculty of Science, University of Zagreb, Bijenička c. 32, HR-10000 Zagreb, Croatia*
[5]*Ecole Polytechnique Fédérale de Lausanne, LPMC, CH-1015 Lausanne, Switzerland*

**Abstract**

The controversy regarding the precise nature of the high temperature phase of $1T$-TiSe$_2$ lasts for decades. It has intensified in recent times when new evidence for the excitonic origin of the low-temperature charge-density wave state started to unveil. Here we address the problem of the high-temperature phase through precise measurements and detailed analysis of the optical response of $1T$-TiSe$_2$ single crystals. The separate responses of electron and hole subsystems are identified and followed in temperature. We show that neither semiconductor nor semimetal pictures can be applied in their generic forms as the scattering for both types of carriers is in the vicinity of the Ioffe-Regel limit with decay rates being comparable to or larger than the offsets of band extrema. The nonmetallic temperature dependence of transport properties comes from the anomalous temperature dependence of scattering rates. Near the transition temperature the heavy electrons and the light holes contribute equally to the conductivity; this surprising coincidence is regarded as the consequence of dominant intervalley scattering that precedes the transition. The low-frequency peak in the optical spectra is identified and attributed to the critical softening of the $L$-point collective mode.

# I. INTRODUCTION

In the continuing search for new collective electronic states, layered materials repeatedly appear as valuable sources and inspiration.[1-5] Transition-metal dichalcogenides (TMDs) have been among the earliest families of quasi-two-dimensional (2D) materials to host the quest, and they continue to provide surprises and puzzles.[1,3,6-8] $1T$-TiSe$_2$ was one of the first TMDs from the IV$b$ group where the charge-density wave (CDW) ordering was identified, yet the mechanism behind the transition has been debated ever since.[1,9-11] Front contenders for the explanation of the CDW ordering in $1T$-TiSe$_2$ are the long-sought "excitonic insulator" and the indirect band-Jahn-Teller mechanism,[1,3,12-16] with the relative importance of electron-electron and electron-phonon interactions still being disputed.[14,17-20] The whole development has further emphasized the need to understand the nature of the high-temperature state that gives birth to the CDW phase. From early to recent times, the question has been reoccurring if this state is a semimetal or a semiconductor.[14,21-24] The reports range from a semiconductor with a finite indirect gap up to 150 meV[14,22] to a semimetal with an indirect band overlap up to 120 meV.[23,24] Understandably, a firm grasp of the high-temperature state is required for the proper understanding of the mechanism of the CDW transition. Two opposing pictures relate to two very different views on transport in the high-temperature phase: The observed nonmetallic temperature dependence of resistivity may be regarded as the consequence of the variation of the carrier density with temperature or as the result of the anomalous temperature dependence of the scattering rate. The two mechanisms also reflect very differently in the temperature dependence of the Hall coefficient, where the compensation of electron and hole contributions opens an additional window into the process.

This paper aims to answer some of these pressing questions in $1T$-TiSe$_2$. Our investigation is based on measuring and analyzing the optical response of the material at the temperatures above the CDW transition.[25] We combine the result of optical and DC transport measurements to identify the contributions of electron and hole subsystems to the low-frequency response and to the transport properties by resolving their respective spectral weights, scattering rates, and mobilities. We find that the carrier density is quasiconstant in temperature, whereas the scattering rates for electrons and holes change in temperature, going beyond the Ioffe-Regel limit. The energy scale $\hbar/\tau$ related to scattering is of the order on the absolute values of the gap (or indirect band overlap) previously quoted, rendering the question of precise bare bands offset less relevant. Instead, the strong scattering regime points to the distinct possibility of dynamically shaped band edges.

## II. THE MATERIAL AND EXPERIMENTAL DETAILS

1$T$-TiSe$_2$ is a quasi-2D layered material where each layer consists of one Ti plane sandwiched between two Se planes. The 1$T$-polytype features one Ti atom in an octahedral arrangement between six Se atoms. In the high-temperature phase the material has a hexagonal structure. All theoretical and experimental papers, including electronic structure calculations and angle resolved photoemission spectroscopy (ARPES) experiments, agree that the conduction-band minimum (at the $L$ point in the Brillouin zone) and the valence band maximum (at the $\Gamma$ point) are close in energy.[14,22,24-27] These two bands have very different dispersions near the Fermi level, with the effective masses as measured in ARPES and concurred by density functional theory calculations, being an order of magnitude bigger in the conduction band than in the valence band. Around 200 K ($T_{CDW}$) the material undergoes the second-order phase transition to the commensurate charge density wave state.[9] The rise of the 2 × 2 × 2 superstructure involves the softening of a phonon at the high-symmetry $L$-point of the Brillouin zone.[28-30] The electrical resistivity $\rho(T)$ exhibits the peak in $d\rho(T)/dT$ around $T_{CDW}$ associated with the superlattice formation, followed by the maximum in $\rho(T)$ near 165 K.[9,10,31] At room temperature the Hall coefficient $R_H$ is positive, indicating that holes are more mobile than electrons in the high temperature phase, but falls to zero near the transition temperature, indicating complete compensation of electron and hole contributions to the Hall signal.[10]

The single crystal 1$T$-TiSe$_2$ samples used in this study were grown by a conventional chemical vapor transport method. The typical size of the crystal samples was 3×2×0.5 mm$^3$. We measured the optical reflectivity (planar response) of 1$T$-TiSe$_2$ in the temperature range between 200 K and 290 K, and a broad frequency range from 40 cm$^{-1}$ to 37 000 cm$^{-1}$. Bruker IFS 113v was used for the far-infrared measurements (40 cm$^{-1}$ – 800 cm$^{-1}$), gold was evaporated onto the sample *in situ*, and reference measurements were performed at each measured temperature. The mid-infrared range was measured with a Bruker Vertex 80v equipped with a HYPERION IR microscope, and freshly evaporated aluminum mirrors were used as a reference. A Woolam variable-angle spectroscopic ellipsometer was used for the near-infrared to ultraviolet frequency range. The measurements were performed in vacuum. This procedure greatly increases the precision of the low-frequency measurements and the outcome of the Kramers-Kronig (KK) transformation, particularly regarding the relative changes that occur with the change in temperature. The measurements were performed at 200 K, 225 K, 250 K and 290 K.

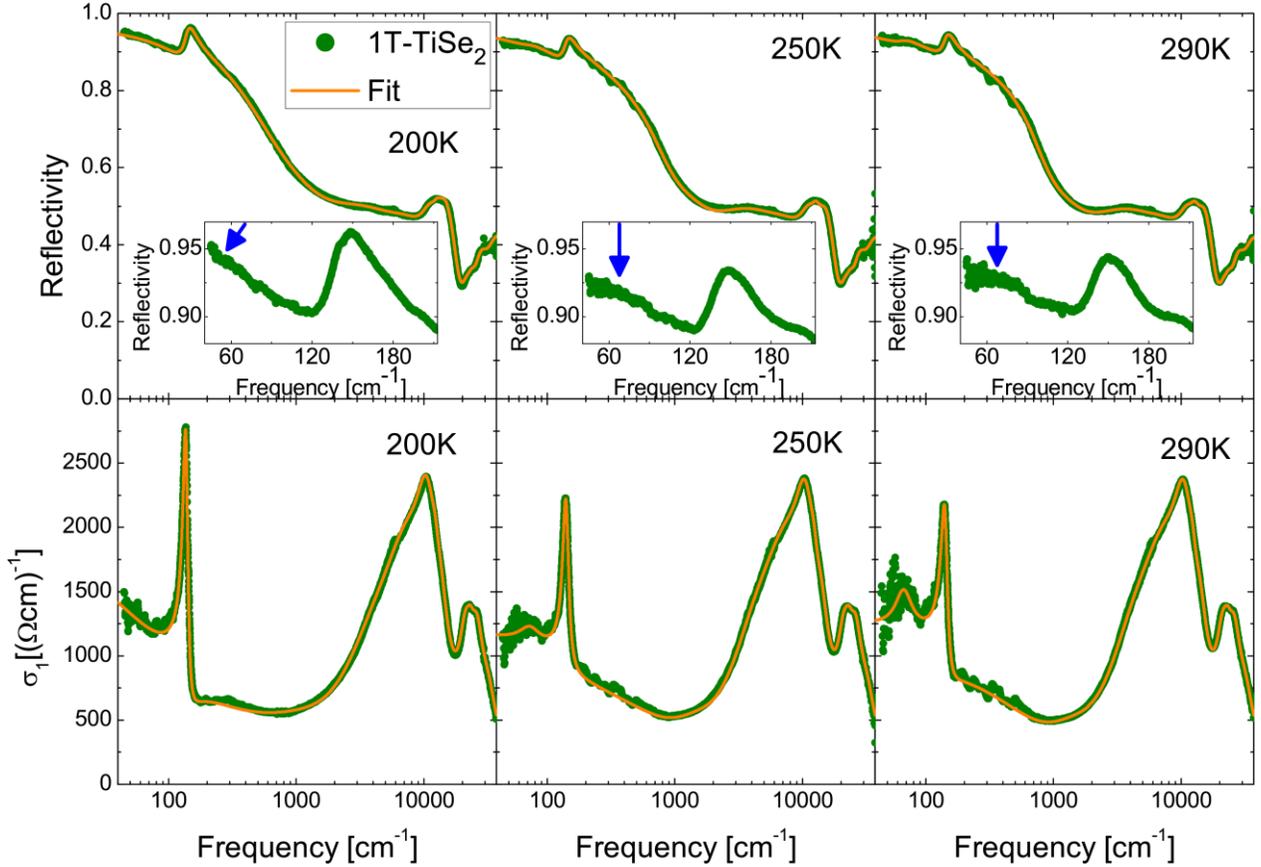

FIG. 1. Experimental data for reflectivity and optical conductivity of $1T$-TiSe$_2$ in the high-temperature phase, 200 K $\leq T \leq$ 290 K in the frequency window from 40 cm$^{-1}$ to 37 000 cm$^{-1}$ (note the logarithmic scale on the horizontal axis). The full lines on top of the data represent the Kramers-Kronig-consistent fits as elaborated in the text. The insets in the upper panels show the reflectivity below 200 cm$^{-1}$. The arrows in insets point to the low-frequency mode discussed in Section III (the arrow points to the frequency $\omega_{LF}$ on the horizontal axis).

## III. EXPERIMENTAL RESULTS AND ANALYSIS

The experimental results for reflectivity $R(\omega)$ are shown in Fig. 1, along with the real part of optical conductivity $\sigma_1(\omega)$, as obtained using the KK transformation.[32-34] Figure 1 also shows the curves that are the results of the modeling to be discussed below. The minimum in optical conductivity occurs around 1000 cm$^{-1}$. This frequency roughly separates the parts of the spectra dominated by contributions of intra- and interband processes.

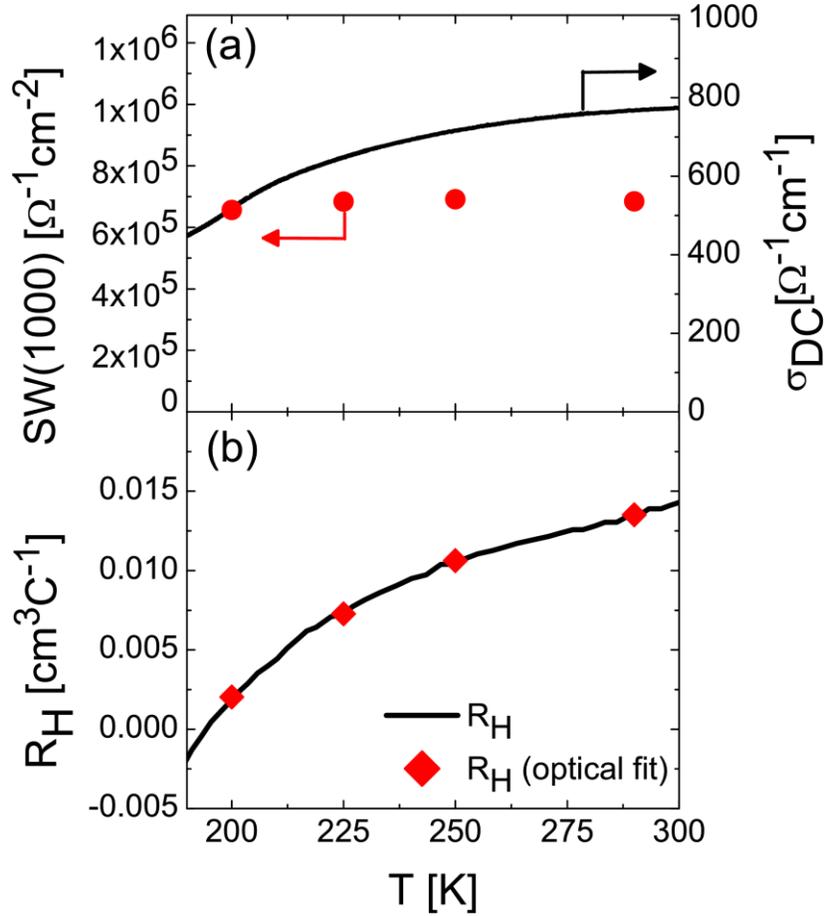

FIG. 2. (a) The comparison of the temperature dependence of DC conductivity $\sigma_{DC}(T)$ and the partial spectral weight $SW(1000) \equiv \int_a^b \sigma_1(\omega) d\omega$, $a = 40$ cm$^{-1}$, $b = 1000$ cm$^{-1}$. The non-metallic increase in conductivity with increasing temperature is not accompanied by any comparable change in the spectral weight. (b) Similarly, the temperature dependence of the Hall coefficient does not reflect the temperature dependence of the carrier density in a generic stoichiometric semiconductor. The figure shows the points that correspond to optical fits in Fig. 1. on the top of the measured Hall coefficient from Ref. 10.

There are several hints about the non-trivial nature of the high-temperature phase that may be obtained already at the level of raw data. The first hint comes from Fig. 2 which shows the temperature dependence of the partial spectral weight, $SW(1000) \equiv \int_a^b \sigma_1(\omega) d\omega$, where $a = 40\,\text{cm}^{-1}$, $b = 1000\,\text{cm}^{-1}$, related to the frequency window that extends from the lowest measured frequency up to the crossover at 1000 cm$^{-1}$. This spectral weight shows no significant variation in temperature. This supports the viewpoint, expressed in some previous papers, of the semi-metallic nature of 1$T$-TiSe$_2$.[23,24] Accordingly, the sizable nonmetallic change in conductivity in the same temperature range may not be ascribed to the variation of the carrier density in temperature, usual for semiconductors.

The second hint comes from the value of the DC conductivity also shown in Fig. 2. Using the value $d_\perp \approx 6\,\text{Å}$ for the separation between TiSe$_2$ layers, one calculates the conductance per sheet, $\sigma_{DC} d_\perp$. This turns to be on the order of conductance quantum, $G_0 = 2e^2/h \approx 7.75 \times 10^{-5}\,\Omega^{-1}$. Meeting this limit in a two-dimensional metal ordinarily means that the strong scattering limit is reached, with the mean free path being on the order of inverse Fermi wave vector, $\ell \sim k_F^{-1}$. For a quasi-two-dimensional semimetal with a finite overlap of conduction and valence bands this also implies that all states in the energy range of band overlap are strongly affected by scattering. The latter point will be further detailed in Section III.

These hints being noted, it may be rightly claimed that 1$T$-TiSe$_2$ is not a simple 2D metal, whereas the quantity SW(1000) shown in Fig. 2 is based on a sharp and somewhat arbitrary cutoff frequency. A meticulous analysis of the spectra is called upon, and this is what we aim to provide in the following paragraphs.

This analysis must also observe the multicarrier nature of transport in 1$T$-TiSe$_2$, witnessed by the Hall coefficient, shown in Fig. 2(b). The temperature dependence of the Hall coefficient as noted previously in Ref. 10 primarily reflects the dependence of mobility on temperature, and the fact that mobilities of electrons ($\mu_e$) and holes ($\mu_h$) get equal near the transition point. This observation is based on the fact that concentrations of electrons and holes are equal in the stoichiometric compound, $n_h = n_e = n$, simplifying the textbook formula for the Hall coefficient,[10,35,36]

$$R_H = \frac{1}{e} \frac{n_h \mu_h^2 - n_e \mu_e^2}{(n_h \mu_h + n_e \mu_e)^2} = \frac{1}{ne} \frac{1 - \frac{\mu_e}{\mu_h}}{1 + \frac{\mu_e}{\mu_h}}. \quad (1)$$

Having comparable mobilities and same concentrations for holes and electrons also implies that their contributions to DC conductivity $\sigma_{DC,h}$ and $\sigma_{DC,e}$ are comparable and get precisely equal in the proximity of the transition point where $R_H$ vanishes.[10] This may appear as a surprising coincidence, given the order of magnitude difference in effective masses found for the two bands. In other words, the scattering rate for lighter carriers (holes) must be an order of magnitude bigger than the scattering rate of heavier carriers (electrons). This great difference in scattering rates for two channels that contribute comparably in the DC limit is expected to show in optical data as well. Indeed, several authors have already noted that the low-frequency electronic response may not be described by a single Drude term.[23,37] It may be expected that several Drude terms are required in the minimal model for the low-frequency optical response in 1$T$-TiSe$_2$ as in some other multiband systems.[38-40]

The modeling of the optical response, to be presented below, is always performed in the KK-consistent manner, i.e. by validating the model against reflectivity and conductivity data sets. The overall fits that emerge are shown in Fig. 1 as full lines on the top of the data.[41] The straightforward approach to modeling of complex optical conductivity, $\sigma(\omega) = \sigma_1(\omega) + i\sigma_2(\omega)$, proceeds by separating the contributions of electrons and phonons $\sigma(\omega) = \sigma_{el}(\omega) + \sigma_{phon}(\omega)$. It is convenient to start with phonons since this part of the spectrum is very simple in the high-temperature phase of 1$T$-TiSe$_2$, consisting of a single infrared-active phonon mode ("Se-mode") at a frequency somewhat above 130 cm$^{-1}$.[23,42,43] Also, compared to all other signals, this mode is very narrow in a frequency, which makes easier to parameterize and separate its contribution from the others. This separation is easy to perform despite pronounced Fano-type asymmetry which signals the coupling of the phonon to the electronic continuum. The well-known Fano formula, $\sigma_{\text{phon}}(\omega) = \varepsilon_0 \omega S_0^2 \left(1 - \frac{i}{q_0}\right)^2 \left(i(\omega_0^2 - \omega^2) + \omega\gamma_0\right)^{-1}$ is used to parameterize the contribution.[44-47] Figure 3 illustrates how it appears in the full spectrum decomposition at 250 K.

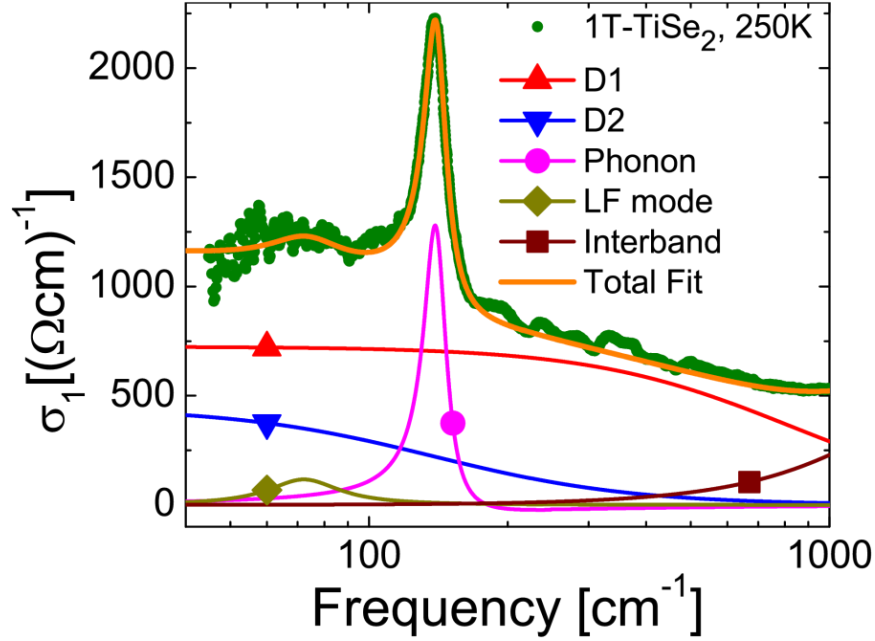

FIG. 3. The real part of the optical conductivity of 1T-TiSe$_2$ below 1000 cm$^{-1}$ at 250 K, along with the decomposition into components as discussed in the text. The related decomposition for the imaginary part is provided in Supplemental Material.[48]

As usual, the electronic part is parameterized by the Drude-Lorentz model,

$$\frac{\sigma_{el}(\omega)}{\varepsilon_0} = \sum_j \frac{\omega_{p,Dj}^2}{\gamma_{Dj} - i\omega} + \sum_j \frac{\omega S_{Lj}^2}{i(\omega_{Lj}^2 - \omega^2) + \omega\gamma_{Lj}}, \quad (2)$$

with multiple Drude and Lorentz terms.

First we address the inter-band excitations that dominate the part of the spectrum above 1000 cm$^{-1}$. Figure 1 shows precise fits to this part of the spectrum produced by using appropriate Lorentz terms.[49] As illustrated in Fig. 2, these terms also produce a weak and featureless tail in the frequency window below 1000 cm$^{-1}$, which is the frequency range of our primary interest.

Thus, apart from simulating the strength, the form, and the KK-consistent analytic structure of the inter-band contributions above 1000 cm$^{-1}$, these Lorentz terms also serve to improve our analysis in the frequency region below 1000 cm$^{-1}$. We also find that the temperature dependence of the interband contribution is small and negligible in comparison to the temperature dependence of the other parts of the spectrum.[49]

At last, we turn to the frequency range below 1000 cm$^{-1}$. We find that this part of the electronic response is very well described by the sum of two Drude contributions, parameterized by very different sets of parameters $(\omega_{p,D_1}, \gamma_{D_1})$ and $(\omega_{p,D_2}, \gamma_{D_2})$. Motivated by previous measurements of electronic spectra and band structure calculations we identify the two terms as responses of holes and electrons ($D_1 \equiv h, D_2 \equiv e$). These parameters are refined not only by producing the best possible fits to the optical data in Fig. 1, but also by obeying the constraints that arise from DC measurements. First, we take care that relative change in optical conductivity in the $\omega \to 0$ limit $\sigma_1(\omega \to 0) = \varepsilon_0 \omega_{p,e}^2 / \gamma_e + \varepsilon_0 \omega_{p,h}^2 / \gamma_h$ follows the relative change in DC conductivity $\sigma_{DC}(T)$. Second, we demand that the mobility ratio that we extract from these parameters, $\mu_e / \mu_h = \omega_{p,e}^2 \gamma_h / \omega_{p,h}^2 \gamma_e$, is consistent with measured temperature dependence of the Hall coefficient through Eq. (1). This requires that the mobility ratio approaches unity around 200 K, as well as that the temperature dependence of this ratio primarily accounts for $R_H(T)$ in Fig. 2(b). Naturally, one should also consider the temperature dependence of the carrier density $n$ in Eq.(1). However, as already stated in relation to integral SW(1000) in Fig. 2(a), this dependence is not expected to be very pronounced. Thus we take the approach where we start with $n$ being temperature independent and check the results for consistency. The value of $n = 1.4 \times 10^{20}$ cm$^{-3}$ is obtained by optimizing the fits in optical conductivity and reflectivity for all temperatures. This value comes within the factor of 2 of the original rough estimate made in Ref. 10 for the carrier density at room temperature, and no compensation effects are taken into account. The resulting parameters that characterize the optical response of electron and hole subsystems are shown in Fig. 4. Figure 4(a) shows the temperature dependence of spectral weights,[32]

$$\mathrm{SW}_a \equiv \int_0^\infty d\omega\, \sigma_{1,a}(\omega) = \frac{\pi \varepsilon_0 \omega_{p,a}^2}{2} = \frac{\pi n_a e^2}{2 m_a^*} = \frac{\pi n e^2}{2 m_a^*},\ (a = h, e), \qquad (3)$$

proportional to the squares of respective plasma frequencies $\omega_{p,h,e}$, whereas the scattering rates $\gamma_{h,e}$ are shown in Fig. 4(b).[50] The spectral weight for holes ($SW_h$) is weakly temperature dependent, the variation being on the order of error bars, consistent with the assumption of the carrier density $n$ being weakly temperature dependent. The variation of the electronic spectral weight ($SW_e$) is also much smaller than the total spectral weight, although some rise in $SW_e$ may be observed as the temperature is lowered. This rise is possibly related to the increased mixing of two bands upon lowering the temperature to be discussed further below. It may be further noticed that the ratio $SW_h / SW_e$ obtained from our data in Fig. 4 comes very close to the value of $m_e / m_h \approx 9.65$ measured in ARPES at 260 K.[14] Also, the absolute values of effective masses using the spectral weights in Fig. 4 and the carrier density $n$ are close to the values derived from ARPES data.[14]

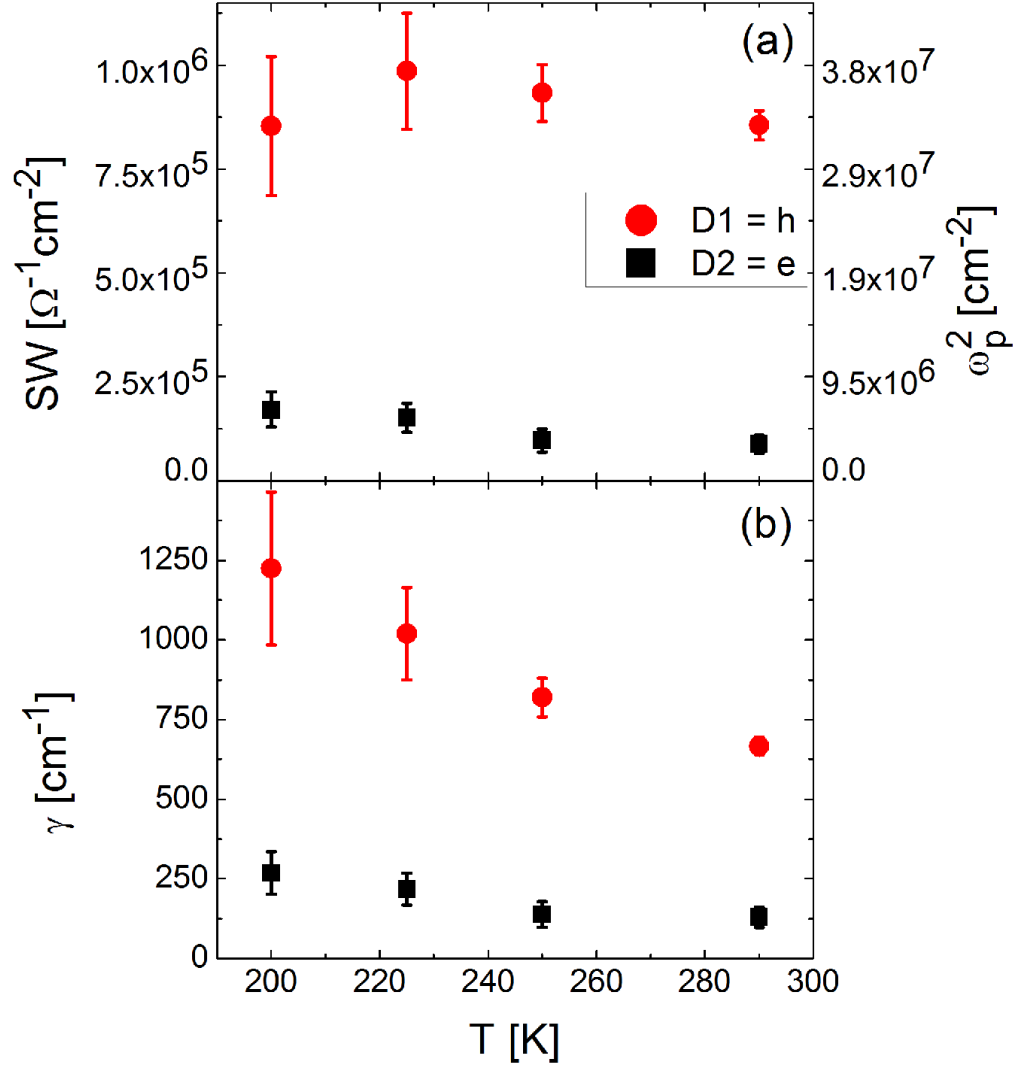

FIG. 4. The results of the Kramers-Kronig-consistent fit to the experimental optical data of 1$T$-TiSe$_2$ in the high-temperature phase (290 K ≥ $T$ ≥ 200 K). (a) Spectral weights of two Drude terms that model the electronic response below 1000 cm$^{-1}$. SW is the spectral weight of the Drude term defined as $\mathrm{SW} = \pi\varepsilon_0\omega_p^2/2$, where $\omega_p$ is the plasma frequency. The right axis shows the square of the plasma frequency. (b) The quantity $\gamma$ is the width (or damping) of the Drude term expressed in cm$^{-1}$, inversely proportional to the transport scattering time.

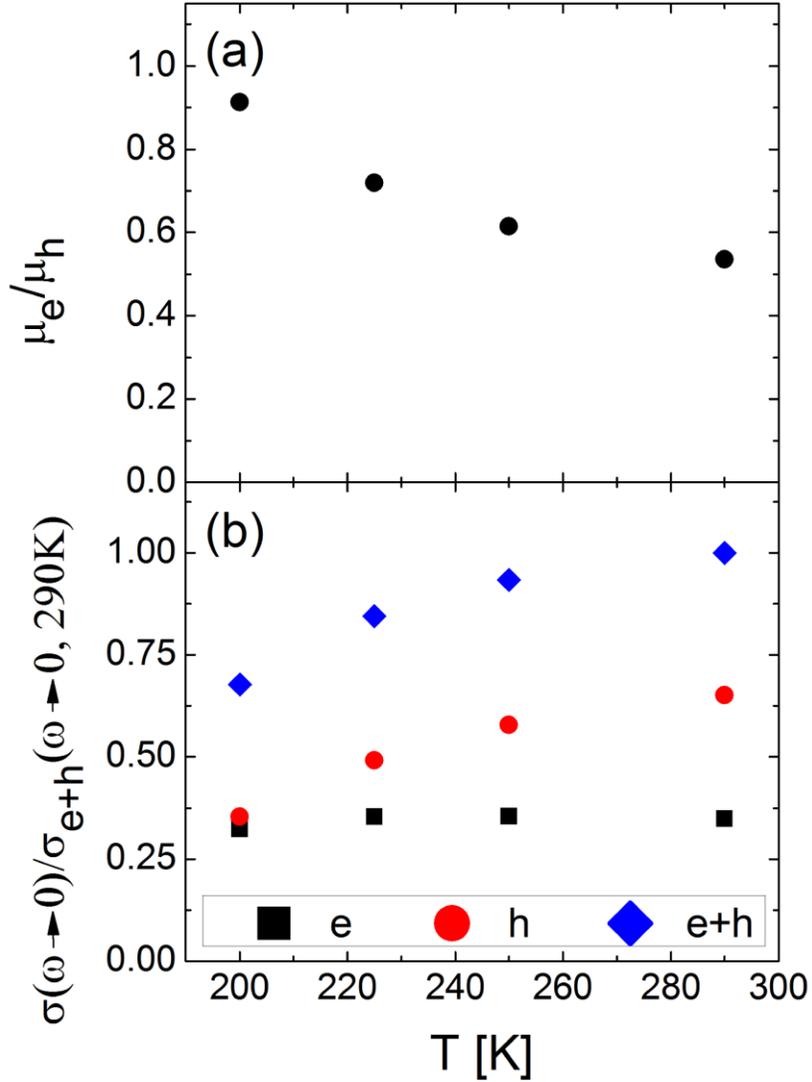

FIG. 5. (a) The temperature dependence of the ratio of electron and hole mobilities as inferred from optical conductivity below 1000 cm$^{-1}$. (b) Contributions of electrons and holes to conductivity in the DC limit.

The temperature dependence of the mobility ratio is shown in Fig. 5(a). The electrons are less mobile than holes at room temperature, but $\mu_e/\mu_h$ increases upon lowering the temperature and approaches unity around 200 K. Figure 5(b) gives the temperature dependence of electron and hole conductivities in the zero frequency limit, $\sigma_{h,e}(\omega \to 0) = (2/\pi)(SW_{h,e}/\gamma_{h,e})$.

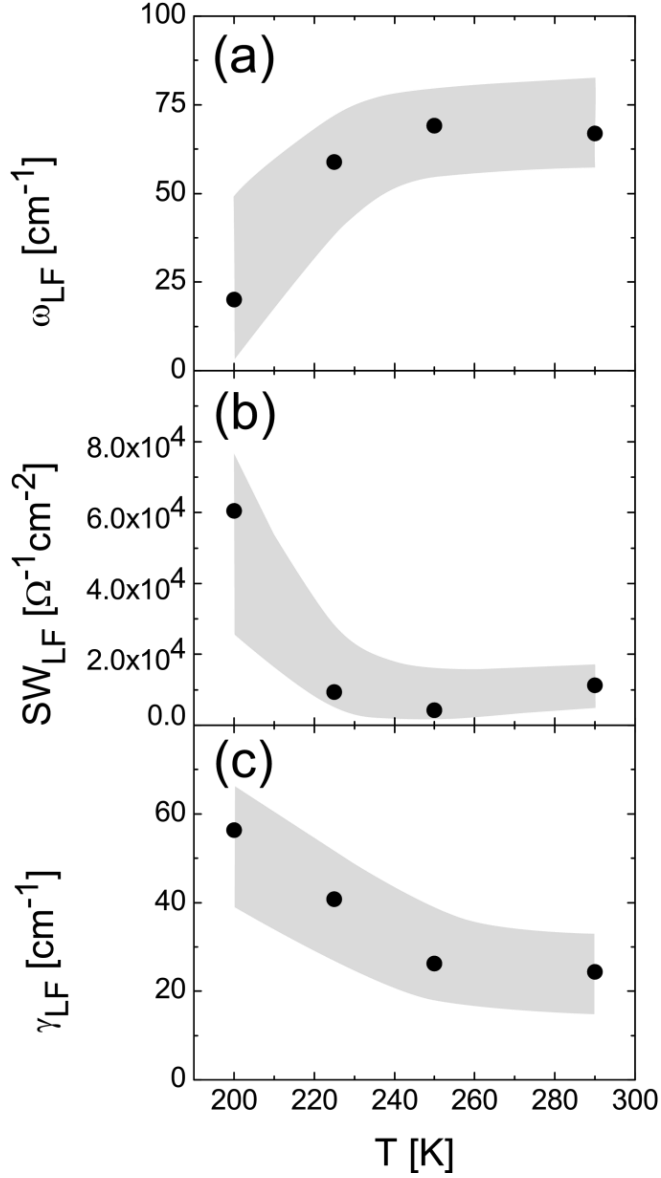

FIG. 6. Temperature dependence of the parameters of the low-frequency mode. As the temperature is lowered towards the transition temperature, the frequency of the mode lowers, whereas the damping and the spectral weight of the mode increase. The gray-shaded areas represent the confidence intervals for the parameters of the low-frequency mode, whereas the points represent the values of parameters used in Figs. 1 and 2.

Finally, we discuss the least pronounced feature of our low-frequency spectra, the low-frequency peak whose spectral weight is much smaller than the spectral weights of the Drude terms. The peak has a characteristic finite frequency ($\omega_{LF} \approx 69$ cm$^{-1}$ at 250 K), which is well below the Se-phonon mode frequency. The peak is considerably broader than the phonon mode, but appears clearly visible as the step in the raw reflectivity data (insets in top panels in Fig. 1). It is also visible in the optical conductivity, despite the experimental noise being very much amplified by the KK transformation (which is typical whenever reflectivity approaches unity). The low-frequency peak is parameterized by the Lorentz form with the parameters shown in Fig. 6.[51] As a consequence of the noise, the KK-consistent fits yield significant error bars for the parameters. However, several features can be followed despite the noise and even observed in raw data: As the transition temperature is approached, the characteristic frequency of the low-frequency mode $\omega_{LF}$ decreases, the spectral weight of the mode increases, and the damping $\gamma_{LF}$ gets larger. Given the values and the temperature dependence of the parameters, it is clear that the observed mode cannot be related to any of the $q=0$ phonon modes in the high-temperature phase of $1T$-TiSe$_2$. On the other hand, the frequency range and temperature dependence of the frequency $\omega_{LF}(T)$ resembles the softening of the phonon branch around the $L$ point in the Brillouin zone (hereafter the "L-phonon mode"), observed through diffuse x-ray scattering above transition temperature.[28-30] However, the phonon at the boundary of the Brillouin zone can be excited by light only if assisted by another excitation which provides the required crystal momentum. This excitation is easy to point to in $1T$-TiSe$_2$, as the very appearance of the Kohn anomaly reflects the coupling of the $L$ point phonon mode to the electronic interband excitations, involving the valence-band states near the $\Gamma$ point and conduction-band states near the $L$ point. This mixing of bands has been previously observed in ARPES at 260 K.[14] The process is illustrated in Fig. 7. In particular, the process of creation of a soft phonon by photons, assisted by the electronic quasielastic interband excitation, is depicted in Fig.7(c). Alternatively, this process may be considered as (soft-)phonon-assisted interband excitation. Even more correctly, it should be regarded as the collective excitation in which phonon and electron-hole excitations are mixed and probably further amplified by the Coulomb (excitonic) correlations in the electron-hole channel.[19,21,52]

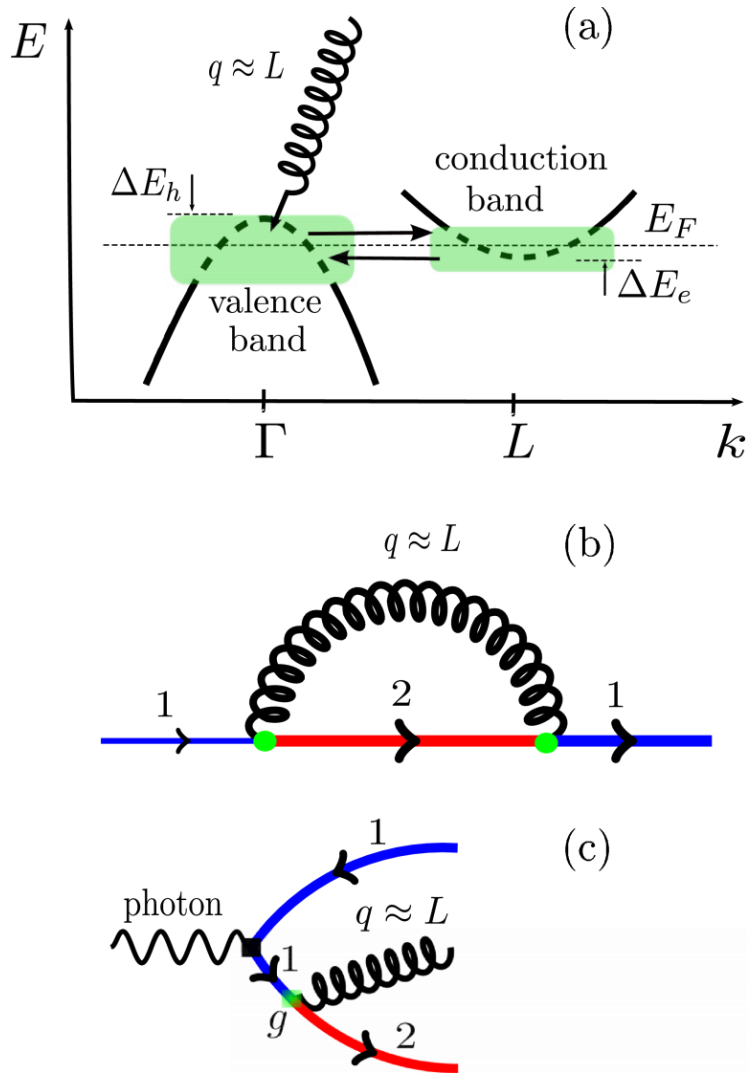

FIG. 7. (a) The schematic of the quasielastic interband scattering that involves the collective soft phonon/boson mode (coiled line) at $q \approx L$. The scattering strongly affects the electronic states near the band edges (shaded areas), which dominantly contribute to low-frequency optical response. (b) The self-energy diagram representing the same scattering process (lowest order of perturbation) related to Eq. (6). The indices 1 and 2 stand for states in conduction and valence bands or vice versa; (c) The process of photon absorption that leads to the generation of soft phonon/boson in the high temperature phase related to the observed low-frequency mode. The process is assisted by the quasi-elastic inter-band electron excitation.

## IV. DISCUSSION

The picture of the semi-metal carried so far may be examined further by comparing all the energy scales involved. First, the measured spectral weight $SW_h$ can be used to assess the energy span $\Delta E_h$ of states in the valence band required to host the holes, in the absence of thermal smearing,

$$\Delta E_h = \frac{\hbar^2 k_{F,h}^2}{2m_h^*} = \frac{\hbar^2 \pi n_h d_\perp}{m_h^*} = \frac{2\hbar^2}{e^2} d_\perp SW_h, \qquad (4)$$

where $k_{F,h}$ and $d_\perp$ denote the Fermi wave vector of the quasi-two-dimensional hole gas, and the lattice constant in the direction perpendicular to the planes, respectively.[46] The value of $\Delta E_h / k_B$ thus obtained from data in Fig. 4(a) lies between 600 K and 700 K, roughly three times the transition temperature. The same is not true for electrons. For similarly defined $\Delta E_e$ the value obtained is approximately tenfold smaller, which implies that electrons in conduction band are in the state of Drude gas, $\Delta E_e \ll k_B T$.[53] The total overlap of the conduction and valence bands may be estimated to (see Fig. 7) $\Delta E \approx \Delta E_h + \Delta E_e \approx \Delta E_h \approx 0.056(\pm 0.005)$ eV ($\Delta E / k_B \sim 650$ K).

On the other hand, the data for the scattering rates, shown in Fig. 4(b), yield $\hbar \gamma_h / k_B$ ranging from 960 K at room temperature to 1760 K around the transition temperature. The relation between bandwidth and scattering rate $\Delta E_h \leq \hbar \gamma_h$ reflects the strong scattering regime announced in the Introduction, here quantified for the hole channel. A similar relation also holds for electrons in the conduction band since $(\Delta E_e / \gamma_e)/(\Delta E_h / \gamma_h) = \mu_e / \mu_h$ is of the order of unity. The overall relation that connects the six energy scales in 1T-TiSe$_2$ in the observed temperature range is

$$\Delta E_e \leq k_B T \sim \hbar \gamma_e \ll \Delta E_h \approx \Delta E \leq \hbar \gamma_h. \qquad (5)$$

Obviously, strong carrier scattering, above the Ioffe-Regel limit $\Delta E_h \sim \hbar \gamma_h$, $\Delta E_e \sim \hbar \gamma_e$, is the essential property of the high-temperature phase of 1T-TiSe$_2$. It is possible that this scattering appears as a wide precursor of the CDW phase, involving the mode at $q \approx L$ that softens throughout the temperature range addressed here.[21,28] Figures 7(a) and 7(b) which illustrate the process also suggest the mechanism in which this scattering acts to equalize the mobilities of carriers in these two bands: The carriers in the valence band are scattered into the conduction band, and vice versa. Of course, in both cases the scattering rates are determined by the density of final states. Therefore, the scattering rate of the lighter carrier type is proportional to the band mass of the heavier one, and vice versa, $\gamma_h \propto g^2 m_e$, and $\gamma_e \propto g^2 m_h$ ( $g$ denotes the relevant electron-phonon coupling, pictured as bullet in Figs. 6 (b) and (c)). This leads to two mobilities being comparable,

$$\mu_h \approx A \frac{1/\gamma_h}{m_h} = A' \frac{1}{m_h g^2 m_e} = A \frac{1/\gamma_e}{m_e} \approx \mu_e, \qquad (6)$$

when this type of scattering process becomes dominant. The proportionality factors $A$ and $A'$ that contain some less interesting factors are introduced for convenience.[54] This may explain the observed tendency of two mobilities becoming close in value as the transition point is approached, and the scattering being increased. Thus the compensation that occurs by having the lighter type of carriers heavily scattered and vice versa is likely nonaccidental. It must be emphasized, however, that diagram in Fig. 7(b) and relation (6) represent the interpocket scattering process to its lowest order, whereas the full solution of the (strong) quasistatic scattering problem and related transport is still pending.

In this respect, we would like to add a comment regarding the running "semimetal vs. semiconductor" dichotomy, mostly spanned by optical and ARPES studies. First, it should be emphasized that the large scattering rate found here is not at odds with ARPES data, where the electronic spectra in the high-temperature phase show substantial smearing near the band edges. In fact, it has been proposed recently by the authors involved in ARPES and theoretical modeling that the low-frequency optical response in 1T-TiSe$_2$ is due to carriers that occupy the long spectral tails of the single-particle spectral functions.[21] Our findings confirm that strong scattering is the primary feature of the electronic response and probably the root of the dichotomy. It must be kept in mind, however, that beyond smearing of the quasiparticle states over some energy range $\hbar\gamma$, scattering is likely to reshape the electronic dispersion on the same energy scale. This is best viewed in quasi-one-dimensional Peierls systems which have been intensely studied for decades and where the CDW transition is also announced by the development of the (Kohn) anomaly in the phonon spectrum, often over an extended temperature range. These systems bear a strong resemblance, as well as some important differences with respect to 1T-TiSe$_2$. For comparison, the non-metallic resistivity behavior in some of these systems extends very much above the CDW ordering temperature.[55-57]. The ARPES studies in the same temperature range have revealed largely suppressed density of states at the Fermi level – the "pseudogap" state, sometimes ascribed to the (bi)polaronic redistribution of the quasiparticle spectral weight away from the Fermi level.[56,58,59] Moreover, similar to our findings in 1$T$-TiSe$_2$, the optical studies in the same temperature range find finite "Drude-type" signals with large widths that indicate strong scattering as well as the "collective mode", which resembles our low-frequency mode.[59-62] Conversely, to the ongoing controversy about the ultimate high-temperature starting point in 1$T$-TiSe$_2$, the starting point for the Peierls mechanism is undoubtedly the metallic state with the Fermi level positioned deeply within the band. Despite that, the observed electronic spectra very much deviate from this simple picture in the wide temperature range above the transition, suggesting that strong scattering processes are responsible for the formation of the pseudogap state. The nonperturbative approach required to capture this effect in the strong-scattering limit beyond the noncrossing approximation has been developed for single-band chain systems[63,64] but still awaited for materials like 1$T$-TiSe$_2$.

## 5. CONCLUSIONS AND OUTLOOK

We have presented a comprehensive study of the high-temperature state of 1$T$-TiSe$_2$ in which the optical spectral weights, scattering rates and mobilities for electrons and holes are separately identified and followed in temperature. The study reveals the origin of the nonmetallic temperature dependence of the resistivity in a wide temperature range above the transition. The temperature behavior of the measured spectral weight rules out the simple explanation of this behavior as a consequence of the semiconductor-like state. Our data and our Kramers-Kronig-consistent analysis favor the picture of the high-temperature phase 1$T$-TiSe$_2$ being a semimetal, with the band overlap which is roughly three times the transition temperature. The temperature variation of the hole- and electron-scattering rates is responsible for the observed nonmetallic temperature dependence of DC resistivity as well as for the temperature dependence of the Hall coefficient. The difference in electron- and hole-scattering rates compensates the big difference in respective band masses, leading to comparable hole and electron contributions to the DC conductivity and their complete cancelation in the Hall coefficient in the proximity of the transition temperature. We argue that this behavior is the consequence of the strong interband quasielastic scattering in the CDW precursor regime.

On the other hand, our analysis also shows that both types of carriers are strongly scattered. The scattering energy scale goes beyond the previously quoted values for the gap, pointing to the possibility that band edges are primarily shaped by quasi-elastic-scattering processes. At this point it is of interest to consider the analogy to some strongly coupled Peierls system, which shares several common features with 1$T$-TiSe$_2$, and where strong carrier scattering and the pseudogap state have been documented in a wide temperature range above the transition. All differences between the systems being acknowledged, the strong scattering and dynamical band reconstruction around the Fermi level may be the origin of the ongoing semimetal vs. semiconductor controversy in 1$T$-TiSe$_2$. Along with the previously observed shadow bands in ARPES, the low-frequency mode reported here may be regarded as another special manifestation of this state.

Admittedly, the low-frequency mode deserves more attention in future studies. It would be also interesting to extend the present study to doped materials. The cross-comparison of optical, photoemission and DC transport experiments on crystals from the same batch would be especially desirable.

## ACKNOWLEDGMENTS

The support and encouragement for this research from Dr. A. Smontara is gratefully acknowledged. K. V. is grateful for the hospitality from the 1.Physikalisches Institut, Universität Stuttgart where the optical measurements were performed. Support from Unity through Knowledge Fund (UKF Grant No. 65/10), Croatian Ministry of Science, Education and Sports Grant No. 035-0352826-2848, and the Deutscher Akademischer Austauschdienst (DAAD Grant No. A/11/91433) research grant to K. V. are acknowledged. The work in Lausanne was supported by the Schweizerischer Nationalfonds zur Förderung der Wissenschaftlichen Forschung.

SUPPLEMENTAL MATERIAL

SI. THE PARAMETERS FOR THE PHONON PEAK

Here we address the fitting parameters for the phonon signal in the high temperature phase of 1T-TiSe$_2$. The Se-phonon signal in optical conductivity is parameterized by the formula in the main text, after Eq. (2). The dependence of the parameters on temperature is shown in Fig. S1. The relative changes in frequency and spectral weight are of the order of several percent. The relative variations in width and asymmetry are more pronounced, but their precise values are also more difficult to determine by the fitting procedure.

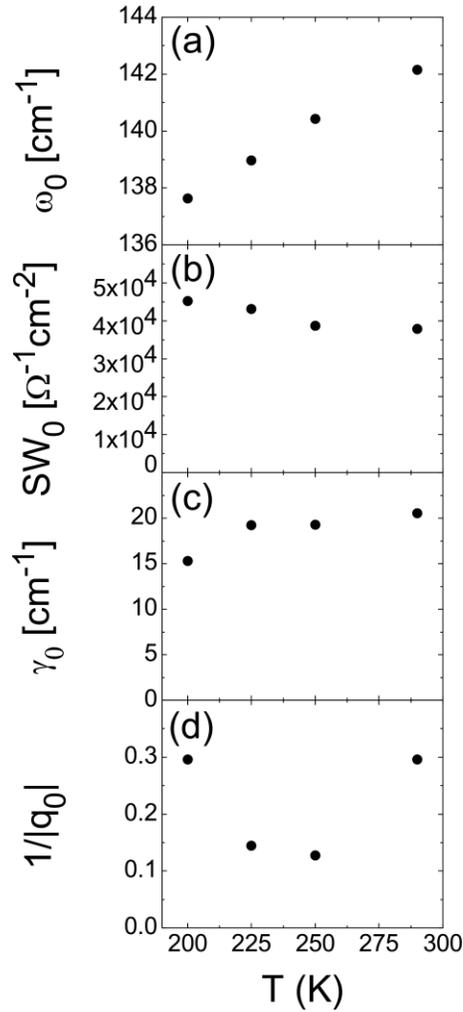

Fig. S1. The dependence of the parameters of the Se-phonon signal on temperature.

## SII. THE INTER-BAND RESPONSE

Here we present the parameterization of the inter-band optical response of 1T-TiSe$_2$ at 250 K. The optical response is modeled by the sum of Lorentzian terms, as specified by Eq. (2). Very precise fit to the measured signal requires multiple terms. The example of fit parameters at 250 K is given in Table I. The significance of the fit is to fix, as precisely as possible, the strength, form, and KK-consistent analytic structure of the inter-band contributions to the optical conductivity/reflexivity, thus improving our analysis in the low frequency domain.

TABLE I. Parameters for all Lorentzian terms used for modeling the inter-band response of 1T-TiSe$_2$ at 250 K.

| No. | $\omega_0$ (cm$^{-1}$) | $\omega_p$ (cm$^{-1}$) | $\gamma$ (cm$^{-1}$) |
|---|---|---|---|
| 1 | 1462 | 4324 | 1785 |
| 2 | 4177 | 15098 | 5409 |
| 3 | 7587 | 26943 | 9066 |
| 4 | 10600 | 17238 | 4882 |
| 5 | 12243 | 4188 | 1924 |
| 6 | 14188 | 9024 | 3992 |
| 7 | 21035 | 13855 | 5738 |
| 8 | 23062 | 6421 | 3814 |
| 9 | 26020 | 16257 | 7231 |
| 10 | 31941 | 19104 | 13626 |

The temperature dependence of the inter-band spectra may be also of interest: We find that the temperature dependence of the inter-band spectra and the influence of its tail to the frequency region below 1000 cm$^{-1}$ is negligible in comparison the temperature dependence produced by other types of excitations. More precisely, we find that the interband contribution to the optical conductivity above roughly 3000 cm$^{-1}$ is for the most part constant in temperature (within 1%). The maximum variation of 4% appears in the rather wide region centered around 17700 cm$^{-1}$. This heightening of the optical conductivity with decreasing temperature, roughly between 900

cm$^{-1}$ and 3000 cm$^{-1}$ there is, as already described in literature [G. Li, et al, Phys. Rev. Lett. 99, 027404 (2007), Ref. [23] in the paper]. The same reference also shows that this heightening develops into a full blown feature in the low temperature CDW phase. However, as already stated, in the high temperature phase the influence of these changes to the low-frequency spectrum below 1000 cm$^{-1}$ is negligible in comparison the temperature dependence produced by other terms. This is quantitatively illustrated in Fig. S2.

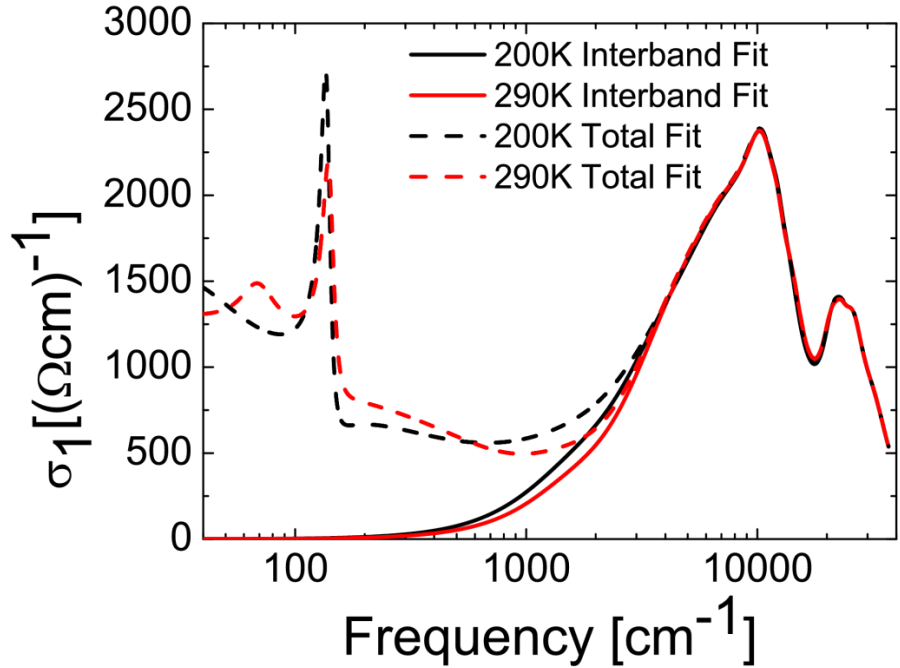

Figure S2. The figure shows our full fits to optical conductivity at 290 K and 200 K (dashed lines). Additional curves shown in solid lines represent the respective inter-band components.

SIII. THE DECOMPOSITION OF THE SPECTRUM BELOW 1000 cm$^{-1}$

Fig.3 in the paper shows the decomposition of the real part of the optical conductivity at 250 K into contributions from various types of excitations. Fig. S3 illustrates how the same decomposition shows in the imaginary part of the optical conductivity.

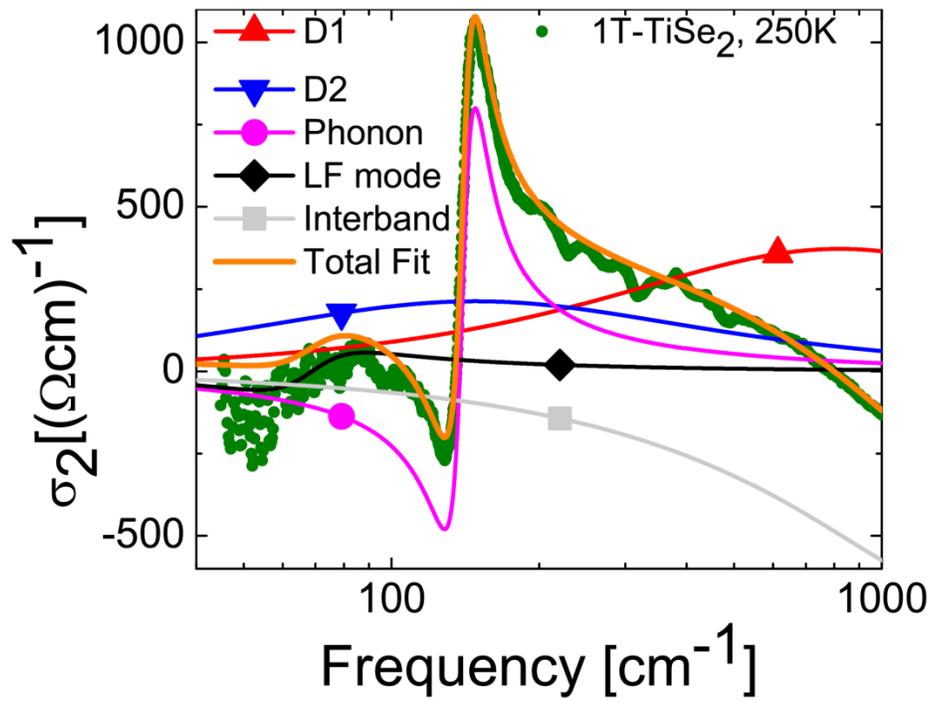

Figure S3. The decomposition of the imaginary part of the optical conductivity below 1000 cm$^{-1}$ at 250 K into contributions from various excitation modes.